\documentstyle[aps,epsf,prl,multicol]{revtex}

\begin{document}
\draft
\title{The quantum Frenkel-Kontorova model: a squeezed state approach}
\author{Bambi Hu$^{[1,2]}$, Baowen Li$^{[1]}$, and Wei-Min Zhang$^{[3]}$}
\address{
$^{[1]}$Department of Physics and Centre for Nonlinear Studies, Hong 
Kong Baptist University, Hong Kong, China\\
$^{[2]}$Department of Physics, University of Houston, Houston TX 77204, USA\\
$^{[3]}$Institute of Physics, Academia Sinica, Taipei, Taiwan 11529} 
\maketitle

\begin{abstract}
The squeezed state is used to study
the one-dimensional quantum mechanical Frenkel Kontorova model. A
set of coupled equations 
for the particle's expectation value and the fluctuations for the 
ground state are derived. It is shown that quantum fluctuations renormalize 
the standard map to an effective sawtooth map. The mechanism underling 
provides an alternative and simple explanation of dynamical localization in 
quantum chaos.
\end{abstract} 

\begin{multicols}{2}

The Frenkel-Kontorova (FK) model describes an atomic chain connected by
harmonic springs subjected to an external sinusoidal potential.  This
model has been widely used to model the crystal dislocations\cite{Nabarro}, 
adsorbed epitaxial monolyaers\cite{Ying}, and incommensurate 
structures\cite{Bak}. The
existence of two competing periodicities may lead to a rich behavior of
the state configurational properties of the particles\cite{Aubry}. 
Recent years has witnessed the application of the FK model
to the study of transmission in Josphson junction and atomic-scale 
friction -nanoscale 
tribology, in which the quantum effects is very essential\cite{tribology}.
 
To get a deep understanding of the nanoscale tribology, it is very
necessary to study the quantum FK model. However, in contrast to the
classical FK model, up to now only a few works have been devoted to the
effects of the quantum fluctuations in the FK 
model\cite{BGS89,BPC94}. 

Like thermal fluctuations in classical systems at finite temperatures,
quantum fluctuations play a very important role in quantum systems with 
finite $\hbar$. In particular, they become crucial and very
important at zero temperature, when  thermal fluctuations vanishe. 
The study of quantum fluctuations becomes an important topic in 
quantum phase transitions \cite{SY95}  and quantum chaos. 

Among many useful tools in study of quantum fluctuations, the squeezed
state, which is a generalization of the coherent state, has been proved to
be very useful in dealing with many-body problems \cite{ZF905,Tsui91}. In
this Letter, we shall study the effect of quantum fluctuations in the
one-dimensional FK model by using the squeezed state approach.  As we
shall see later, a set of coupled equations for the expectation value and
the flucatuation of the particle will be derived for the ground state at
zero temperature. We discover analytically how quantum fluctuations
renormalize the external potential, which leads to the transition of the
standard map in classical FK model to the sawtooth map in the quantum FK
model. The results are found to be in a good agreement with that of
quantum Monte Carlo (QMC) method. 

The Hamiltonian operator of the one-dimensional standard FK model is,
\begin{equation}
\hat{\cal H} = \sum_i \left[\frac{\hat{p}_{i}^{2}}{2m} + 
\frac{\gamma}{2}(\hat{x}_{i+1} - \hat{x}_{i} - a)^2 - 
V\cos(q_0\hat{x}_i)\right]. 
\label{CLHam}
\end{equation}
Here $m$ is the mass of 
particle, $\gamma$ the elastic constant of the spring, $2\pi/q_0$ the 
period of external potential. $V$ is the strength of the external 
potential, $a$ the equilibrium distance 
between two nearst neighbor particles as the external potential vanishes.
For convenience, we can rescale the variables into dimensionless one 
and obtain a new Hamitonian
\begin{equation}
\hat{H} = \sum_i \left[\frac{\hat{P}_{i}^{2}}{2} + 
\frac{1}{2}(\hat{X}_{i+1} - \hat{X}_{i} - \mu)^2 - K\cos(\hat{X}_i)\right].
\label{QMHam}
\end{equation}
where $K=Vq^2_0/\gamma$ is the rescaled strength of external potential.
The effective Planck constant 
$\tilde{\hbar} = \hbar\frac{q_0^2}{\sqrt{m\gamma}},$
is the ratio of the 
natural quantum energy scale ($\hbar\omega_0$) to the natural classical 
energy scale ($\gamma/q_0^2$). $\omega^2_0=\gamma/m$.

The position and momentum operators for the $i$th particle are written as
\begin{equation}\begin{array}l
\hat{X}_i = \sqrt{\frac{\tilde{\hbar}}{2}} ( \hat{a}^{+}_{i} + \hat{a}_i),\\
\hat{P}_i = i \sqrt{\frac{\tilde{\hbar}}{2}} ( \hat{a}^{+}_{i} - \hat{a}_i)
\end{array}
\label{xpoprator}
\end{equation}
Here, $\hat{a}^{+}_{i}$ and $\hat{a}_i$ are boson creation and annihilation 
operators which satisfy the canonical commutation relations: 
$[\hat{a}_i,\hat{a}_j^+]=\delta_{ij}$, $[\hat{a}_i,\hat{a}_j]=0$ and 
$[\hat{a}_i^+,\hat{a}_j^+]=0$.

The squeezed state 
$|\Phi\rangle$ is 
defined by the ordinary harmonic oscillator displacement operator 
$e^{\hat{S}(\alpha)}$ acting on a squeezed vacuum state,
\begin{equation}
|\Phi(\alpha,\beta)\rangle = e^{\hat{S}(\alpha)} e^{\hat{T}(\beta)}|0\rangle,
\label{SS}
\end{equation}
where
\begin{equation}\begin{array}l
\hat{S}(\alpha) = \sum_i(\alpha_i \hat{a}^+_i - \alpha^*_i 
\hat{a}_i),\\
\hat{T}(\beta) = \frac{1}{2}\sum_{ij}(\hat{a}^+_i\beta_{ij} 
\hat{a}^+_j - \hat{a}_i \beta^+_{ij} \hat{a}_j).\end{array}
\label{ST}
\end{equation}
$|0\rangle $ is the vacuum state and $\hat{a}_i|0\rangle= 0$. 
$\hat{S}^{+}(\alpha) 
= - \hat{S}(\alpha), \hat{T}^{+}(\beta) = - \hat{T}(\beta)$.
For simplicity, in what follows we will use the 
abbreviation $|\Phi\rangle = |\Phi(\alpha,\beta)\rangle$.

It must be noted that if we set $\beta =0 $, the squeezed state 
is reduced to the coherent state. As we shall see 
later, the coherent state is not able to allow us to study the 
fluctuations.

Using $|\Phi\rangle$ as a trial wave function for Hamiltonian (\ref{QMHam}),
we can easily 
find the expectation values of the coordinate 
and the momentum operators of the $i$th particle \cite{ZF905},
\begin{equation}\begin{array}l
\bar{X}_i  \equiv  \langle\Phi|\hat{X}_i|\Phi\rangle = 
\sqrt{\frac{\tilde{\hbar}}{2}} (\alpha^*_i + \alpha_i),\\
\bar{P_i}  \equiv  \langle\Phi|\hat{P}_i|\Phi\rangle= -i 
\sqrt{\frac{\tilde{\hbar}}{2}} (\alpha^*_i - \alpha_i).
\end{array}
\label{expect}
\end{equation}
Fluctuations in the coordinate and the momentum are given by
\begin{eqnarray}
\Delta X^2_i  &\equiv  &\langle\Phi|(\hat{X}_i - \bar{X}_i)^2|\Phi\rangle = 
\tilde{\hbar} G_{ii},\nonumber\\
\Delta P^2_i  &\equiv & \langle\Phi|(\hat{P}_i - 
\bar{P}_i)^2|\Phi\rangle,\nonumber\\ &=& 
\tilde{\hbar} (\frac{G^{-1}_{ii}}{4} + 
4\sum_{l,k}\Pi_{il}G_{lk}\Pi_{ki}).
\label{fluct}
\end{eqnarray}
The fluctuation covariance 
between the $i$th particle and the $j$th particle is
\begin{equation}
\Delta X_i \Delta X_j \equiv \langle\Phi|(\hat{X}_i - \bar{X}_i)(\hat{X}_j - 
\bar{X}_j)|\Phi\rangle = \tilde{\hbar} G_{ij},
\label{covar1}
\end{equation}
where, $G_{ij}$ and  $\Pi_{ij}$  are
\begin{equation}\begin{array}l
G_{ij}  =  \frac{1}{2} (\cosh^2\sqrt{\beta\beta^+} +
\sinh^2\sqrt{\beta\beta^+})_{ij}
+ \frac{1}{2}(M\beta+ \beta^+M)_{ij},\\
\Pi_{ij} = \frac{i}{4}G^{-1}_{ij} (M\beta-\beta^+M)_{ij},
\end{array}
\label{GPI}
\end{equation}
where,
\begin{equation}
M= 
\frac{\sinh\sqrt{\beta\beta^+}\cosh\sqrt{\beta\beta^+}}{\sqrt{\beta\beta^+}}.
\label{Matrix}
\end{equation}
Since $\beta$ is a symmetric matrix, $G_{ij} = G_{ji}$ and  
$\Pi_{ij} = \Pi_{ji}$. 
Furthermore, using the following very important relation,
\begin{equation}
\langle\Phi|\cos\hat{X}_i|\Phi\rangle 
=\exp\left(-\frac{\tilde{\hbar}}{2}G_{ii}\right)  \cos\bar{X}_i,
\label{cos}
\end{equation}
we can finally obtain the expectation value of the Hamiltonian $\hat{H}$,

\begin{eqnarray}
\bar{H} &\equiv & \langle\Phi|\hat{H}|\Phi\rangle\nonumber\\
&=&\sum_i
\frac{1}{2}\left(\bar{P}_i^2 + \tilde{\hbar}(\frac{G^{-1}_{ii}}{4} + 
4 \sum_{l,k}\Pi_{il}G_{lk}\Pi_{ki})\right)\nonumber\\
& + & \sum_i \frac{1}{2} (\bar{X}_{i+1} -\bar{X}_i - \mu)^2\nonumber\\
& + & \sum_i \frac{1}{2} \left(
\tilde{\hbar} (G_{ii} + G_{i+1 i+1}) - 2\tilde{\hbar} G_{i+1 i} 
\right)\nonumber\\ 
& - & \sum_i K\exp\left(-\frac{\tilde{\hbar}}{2}G_{ii}\right)  
\cos\bar{X}_i .
\label{ExpcHam}
\end{eqnarray}

It is worth noting that the variables $\bar{X}_i$ and $\bar{P}_i$, and $ 
G_{ij}$ and $\Pi_{ij}$ form explicitly canonical conjugates\cite{Tsui91}. 
To find the 
ground state of the quantum FK model, we shall take a variational approach, 
and these four variables are regarded as variational variables.   
Variation with respect to (w.r.t.) 
$\bar{P}_i$ immediately yields $\bar{P}_i =0$, and w.r.t. $\bar{X}_i$ yields,

\begin{equation}
\bar{X}_{i+1} - 2
\bar{X}_{i} +
\bar{X}_{i-1}  =  K_i \sin\bar{X}_i .
\label{position}
\end{equation}
where $K_i = K \exp\left(-\frac{\tilde{\hbar}}{2}G_{ii}\right)$,
which determines the expectation value of the particle's coordinate. 
Unlike its classical counterpart ($\tilde{\hbar} =0$, 
$K_i = K$), this equation is 
coupled with the quantum fluctuation by $\tilde{\hbar}G_{ii}$.
Variation w.r.t. $\Pi_{ij}$ leads to $4\tilde{\hbar}G_{ij}\Pi_{ji} =0$. 
Since the fluctuation $G_{ij}$ 
cannot be zero, they are always positive, we have $\Pi_{ji} =0$. To obtain 
equation 
for $G_{ij}$, we first take variation w.r.t. $G_{ik}$ and note the 
following relation: $\frac{\delta G_{ij}}{\delta G_{kl}} = \delta_{ik} 
\delta_{jl}$,
where, $\delta_{ik}$ and $\delta_{jl}$ are Dirac delta functions. We then 
multiply both sides of the equation by $G_{kj}$ and take summation over 
$k$. Finally we get the closed equations for the covariance $G_{ij}$,
\begin{equation}
(GF)_{ij} =  G_{i-1 j} + G_{i+1 j},
\label{covar}
\end{equation}
where
\begin{equation}
F_{ij}  =  \delta_{ij} \left( 1 
+ \frac{K_i}{2}\cos\bar{X}_i\right)-\frac{(G^{-2})_{ij}}{8}.
\label{F}
\end{equation}
This is a set of equations determining the quantum fluctuations of 
the particles,
$G =\{G_{ij}\}$. $G$ is a $N\times N$ symmetric matrix which provides all
the fluctuation information. Its diagonal elements give the variance of
each particle, while its off-diagonal elements give the covariance
between particles, from which we can calculate the correlation function of
the quantum fluctuation. These equations are coupled with the
expectation value $\bar{X}_i$. 

Up to this point, we have obtained $N\times (N+1)/2 + N$ equations for 
all variables. These equations provide a qualitative 
picture about the system before we proceede to do any detailed numerical 
analysis.  In fact, if 
we introduce a new variable, $I_{i+1} = \bar{X}_{i+1} - \bar{X}_i$,
Eq.(\ref{position}) can be cast into the  
map,
\begin{equation}\begin{array}l
I_{i+1} = I_i + K_i \sin\bar{X}_i,\\
\bar{X}_{i+1} = I_{i+1} + \bar{X}_i.
\end{array}
\label{Xmap}
\end{equation}
In the same manner, by denoting $Q_{i+1j} = G_{i+1j} -G_{ij}$, 
we can also write Eq.(\ref{covar}) into the form of a map,
\begin{equation}\begin{array}l
Q_{i+1j} = Q_{ij} + (G(F - 2))_{ij},\\
G_{i+1j} = G_{ij} + Q_{i+1j}.
\end{array}
\label{Gmap}
\end{equation}
The difference between the classical 
($\tilde{\hbar}=0$) 
and the quantum FK model from Eq.(\ref{Xmap})  is readily seen. In the 
classical 
case the control parameter, namely the amplitude of the external potential,
does not change with the position index $i$. 
However, in the 
quantum case, due to the quantum fluctuation, the amplitude of 
the effective 
external potential which acts on the particle changes from particle to 
particle. Because  $G_{ii} > 0$ for any nonzero $\tilde{\hbar}$, $K_i<K$, 
which means that the quantum fluctuation reduces the external potential 
strength acting on the particle. Another important difference is that, in 
the classical case, the coordinates of the atoms in the ground state are 
determined by the standard map, whereas in the quantum case 
they are determined by $(N+1)$ coupled two-dimension maps. This 
makes 
the quantum FK model extremly difficult to deal with analytically.

Before we turn to the numerical calculation,  it is worth pointing out 
that in the case of $\beta =0$ in Eq.(\ref{SS}), $G_{ij} 
=1/2$ (for all $i,j=1,2,\cdots,N$),
which is the result of the coherent state theory. It is obvious that this 
cannot be the case for a real quantum FK model. So, coherent state is 
not suitable for the study of quantum FK model. 

We now make some comparisons with the  quantum Monte Carlo (QMC) 
method.
As mentioned before, to find the solution from two sets of equations 
Eqs.(\ref{position}) and (\ref{covar}) is
equivalent to find the periodic orbit in a $2(N+1)$-dimensional map 
Eq.(\ref{Xmap}) and Eq.(\ref{Gmap}). This is still a big problem in 
the nonlinear dynamics to be solved. Nevertheless, we can make a
numerical test of the Eq.(\ref{position}) to see whether this 
equation can give rise to the "sawtooth map"\cite{BGS89}.

In Fig. 1 we show the quantum Monte Carlo results (left
column) and the results calculated from Eq.(\ref{position})
(right column) by using the QMC's fluctuation data $G_{ii}$ in the 
supercritical regime ($K=5$) with
$\tilde{\hbar}=0.2$ for an incommensurate ground state. 
In our quantum Monte Carlo computation, 
as usual, we use the continued fraction expansion for
the golden mean winding number $(\sqrt{5} -1)/2$. Thus, we use $Q$ particles 
which substrated into $P$ external 
potentials with period of $2\pi$. Periodic boundary condition is used: 
\cite{Linote}. $\bar{X}_{Q+i} = \bar{X}_{i} + 
2\pi P $. The winding number is $P/Q$. The results shown in the
figure is for $P/Q=34/55$.

By using QMC calculation, we obtained the expectation value of the atom's 
coordinate, from which we can construct the so called quantum Hull 
function (QHF), namely $\bar{X}_i$ (mode $2\pi$) versus the unperturbed ones 
$2\pi i P/Q$ (mode $2\pi$), which is shown in top-left in Fig. 1. 
The $g$-function which defined by \cite{BGS89}

\begin{equation}
g_i \equiv K^{-1}(\bar{X}_{i+1} - 2\bar{X}_i + \bar{X}_{i-1}),
\label{Gfunc}
\end{equation}
is shown in the middle-left of Fig. 1 from the QMC data. 
The quantum fluctuation $G_{ii}$ calculated from QMC is shown also in the 
bottom-left of Fig. 1.

To compare the squeezed state results with those from QMC, we substitute 
$G_{ii}$ calculated from QMC into Eq.(\ref{position}) and then 
compute the expectation value of the particles' coordinates by using 
Aubry's gradient 
method. We then construct the quantum Hull function and
quantum $g$-function, which is shown in the right column of Fig. 1.

\begin{figure}
\epsfxsize=8cm
\epsfbox{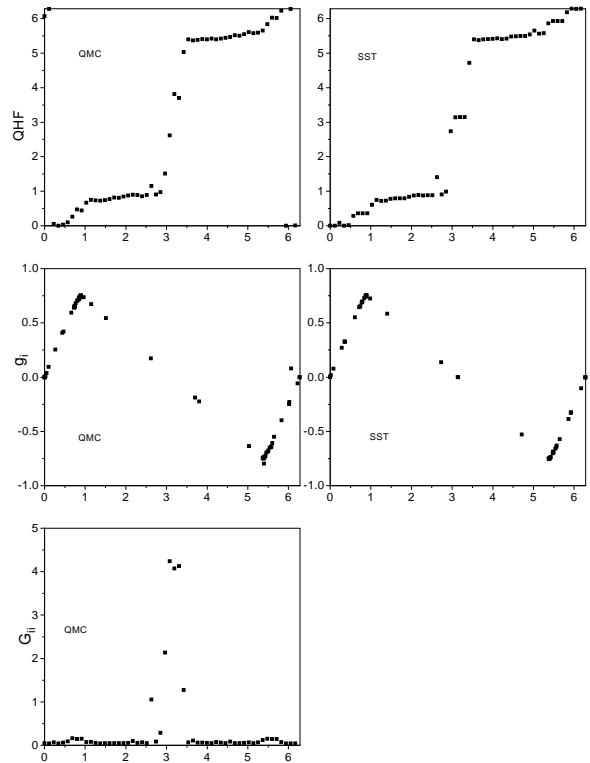}
\narrowtext
\caption
{
Comparision between the quantum Monte Carlo (QMC) results (left 
column) and 
those from the squeezed state theory (SST) Eq.(\ref{position}) (right 
column) in the supercritical regime where $K=5$ with $\tilde{\hbar}=0.2$.
} 
\end{figure}


The results from Eq.(\ref{position}) (right column) agree
surprisingly well with those from QMC for the quantum Hull function as well 
as the $g$-function. The most striking feature to be noted is 
the sawtooth shape of the $g$-function in the supercritical regime.
This phenomenon was 
first observed by Borgonovi {\it et al}\cite{BGS89} in their QMC 
computation and 
has been explained as a tunneling effect. Later on Berman {\it et al} 
\cite{BPC94} recovered this phenomenon by using a mean field 
theory including the contribution from quasidegenerate states.
In the framwork of the squeezed state theory, this quantum 
sawtooth map is just a straightforward result of  
Eq.(\ref{position}), which results from the quantum fluctuations. 

Our result demonstrates that the squeezed state approach 
indeed captures correctly and nicely the effects of quantum 
fluctuations. 

Finally, we would like to point out that the 
mechanism of the reduction of the effective potential due to the 
quantum fluctuations demonstrated above can be applied to explain the 
quantum suppresion of chaos and relevant phenomena such as  
dynamical localization in quantum chaos. 
The dynamical  localization is a well-established fact, it was observed 
by Casati {\it 
et al}\cite{CCFI79} numerically almost 20 years ago and confirmed 
recently in several different experiments 
such as hydrogen atoms in microwave fields and so on\cite{Hatoms}. Its 
underlying mechanism is still not completely understood.
Here, we shall demonstrate that by applying the squeezed state 
approach to the kicked rotator, we could obtain a simple and clear 
picture of the dynamical localization. 

Using the squeezed state,  we obtain a map like 
Eq.(\ref{Xmap}) 
for the expectation value of the angular variable and angular 
momentum. But the equation determining $G_{ii}$ is 
different from Eq.(\ref{covar}), in this case it can be numerically 
calculated. We found that the 
fluctuation $G_{ii}$ grows quadratically with time (kicks), eventually  
the strength of external control parameter $K_i$ becomes very small, thus 
the classical chaos is completely suppressed and leads to the dynamical 
localization. This gives us an alternative explanation and a very simple 
picture of the dynamical localization. In turn, it shows that the 
squeezed state is a very useful tool in study the phenomena related with 
the quantum fluctuations.

In conclusion, we have derived a set of coupled equations determining the 
expectation values of the coordinate and the quantum fluctuations by using 
the squeezed state as a trial wave function. The results from the 
squeezed-state theory agree with those from the quantum Monte Carlo method 
quite well. 
Furthermore, the squeezed-state results give us a very clear understanding
of the 
renormalization of the standard map in the classical case to the effective 
sawtooth map in the quantum case. Moreover, the squeezed state approach 
provides an alternative and a simple picture of the dynamical 
localization observed in many quantum systems.

We would like to thank Dr. L.H. Tang and all members of Centre for
Nonlinear Studies for stimulating discussions. B. Li thanks F.
Borgonovi for reading the manuscript and many helpful discussions. The
work was supported in part by the grants from Hong Kong Research Grants 
Council (RGC) and the Hong Kong Baptist University Faculty Research 
Grants (FRG).

\end{multicols}

\begin{references}
\bibitem{Linote}
This is slightly different from 
that of Borgonovi {\it et al} \cite{BGS89}, there they set $X_0 = 0. X_Q = 
2\pi P$, we have calculated by QMC  for both cases  and found that the 
numerical results are slightly 
different. We use the periodic boundary condition, since  we believe that 
this is a very general case.
\end{references}
\end{document}